\documentclass[a4paper,10pt]{article}
\usepackage[T2A]{fontenc}
\usepackage{amssymb,amsfonts,amsmath,mathtext,cite,enumerate,float,mathrsfs}
\usepackage[utf8]{inputenc}
\usepackage[english]{babel}
\usepackage{graphicx,color}
\usepackage{wrapfig}
\graphicspath{{./}}

\usepackage{epsfig}
\usepackage{epsf}
\usepackage{subfig} 
\usepackage{feynmp} 
\usepackage{pgf,pgfarrows,pgfnodes,pgfautomata,pgfheaps,multirow}
\usepackage{array}
\usepackage{bm}
\usepackage{latexsym}
\usepackage{pifont}

\newcommand{\be}{\begin{equation}}
\newcommand{\ee}{\end{equation}}
\newcommand{\bea}{\begin{eqnarray}}
\newcommand{\eea}{\end{eqnarray}}
\newcommand\nn{\nonumber}

\title{The decay $\tau \to \pi\omega\nu$ in the extended NJL model}

\author{M. K. Volkov\footnote{E-mail address: volkov@theor.jinr.ru}, 
A. B. Arbuzov\footnote{E-mail address: arbuzov@theor.jinr.ru}, 
D. G. Kostunin\footnote{E-mail address: kostunin@theor.jinr.ru}\\
\it Bogoliubov Laboratory of Theoretical Physics, JINR\\ 
\it Dubna, 141980, Russia}

\begin{document}

\maketitle

\begin{abstract}
The decay width $\tau\to\pi\omega\nu$ in the framework of the extended NJL model is calculated.
The contributions of the intermediate vector mesons $\rho(770)$ and $\rho'(1450)$ are taken into account.
The computed partial width and the spectral function of the decay $\tau\to\pi\omega\nu$ are 
in satisfactory agreement with experimental data.
\\

\\


12.39.Fe 	Chiral Lagrangians 
\end{abstract}



\section{Introduction}
The process of the decay $\tau\to\pi\omega\nu$ is intensively investigating from 
experimental~\cite{Edwards:1999fj,Buskulic:1996qs} as well as from 
theoretical~\cite{LopezCastro:1996xh,FloresTlalpa:2007bt,Guo:2008sh} points of view.
In these works phenomenological models with intermediate vector mesons $\rho(770)$, $\rho(1450)$ 
and $\rho(1700)$ were used. 
In all these models arbitrary parameters was introduced and adjusted to fit experimental data .

On the other hand the similar processes 
$e^{-}e^{+} \to \pi^{0}\gamma, \pi^{0}\rho^{0}, \pi^{0}\omega, \pi^{+}\pi^{-}$ in the extended 
Nambu-Jona-Lasinio (NJL) model~\cite{VolkovWeiss, yadPh, VolkovEbertNagy, UFN} 
were described~\cite{Arbuzov:2010xi,Arbuzov:2011zz,Volkov:2012tk,Ahmadov:2011ey}.

Let us note that NJL~\cite{VolkovAn,pepan86, EbertReinhardt, pepan93, VolkovEbertReinhardt, UFN} model allows us to describe the number of tau lepton decays~\cite{VolkovIvanovOsipov3pi,VolkovIvanovOsipovpigamma,Volkov:2012uh}.
In this work the investigation of the $\tau$ decays is continued and the decay $\tau\to\pi\omega\nu$ is calculated in the framework of the extended NJL model which takes into account intermediate 
$\rho(770)$ and $\rho(1450)$ mesons.

\section{The decay $\tau\to\pi\omega\nu$}
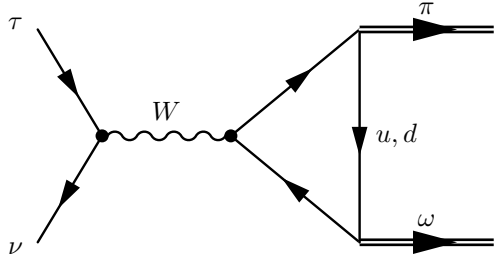
\begin{figure}[ht]
\begin{center}

\begin{fmffile}{contact}

      \begin{fmfgraph*}(200,100)
	      \fmfpen{thin}\fmfleftn{l}{2}\fmfrightn{r}{2}

	      \fmfright{b,a}
	      \fmfleft{f,fb}
	      \fmflabel{$\tau$}{fb}
	      \fmflabel{$\nu$}{f}
	      \fmf{fermion}{fb,v1,f}
 	      \fmf{fermion,tension=.5}{p1,v2,p2}
	      \fmf{boson,lab.side=left,label=$W$}{v1,v2}
 	      \fmf{fermion,lab.side=left,label=$u,,d$}{p2,p1}	      
 	      \fmfdotn{v}{2}

	      \fmf{dbl_plain_arrow,lab.side=left,label=$\pi$}{p2,a}
	      \fmf{dbl_plain_arrow,lab.side=left,label=$\omega$}{p1,b}

	      \fmfforce{120,90}{p2}
	      \fmfforce{120,10}{p1}

	      \fmfforce{170,90}{a}
	      \fmfforce{170,10}{b}

	      \fmfforce{0,90}{fb}
	      \fmfforce{0,10}{f}

      \end{fmfgraph*}

\end{fmffile}
\caption{Contact interaction $W$ boson with a triangle quark loop.}
\label{fig1}
\end{center}
\end{figure}

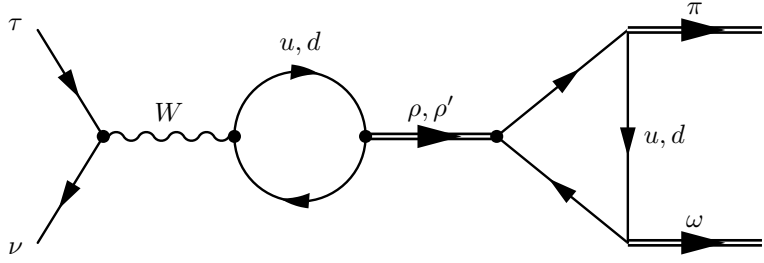
\begin{figure}[ht]
\begin{center}

\begin{fmffile}{rho}

      \begin{fmfgraph*}(200,100)
	      \fmfpen{thin}\fmfleftn{l}{2}\fmfrightn{r}{2}

	      \fmfright{b,a}
	      \fmfleft{f,fb}
	      \fmflabel{$\tau$}{fb}
	      \fmflabel{$\nu$}{f}
	      \fmf{dbl_plain_arrow,lab.side=left,label=$\rho,,\rho'$}{v1,v4}
	      \fmf{boson,label=$W$}{v2,v3}
	      \fmf{fermion,left,tension=.5}{v1,v2}
	      \fmf{fermion,label=$u,,d$,left,tension=.5}{v2,v1}
	      \fmf{fermion}{fb,v3,f}
 	      \fmf{fermion,tension=0.5}{p1,v4,p2}
 	      \fmf{fermion,lab.side=left,label=$u,,d$}{p2,p1}	      
	      \fmfdotn{v}{4}

	      \fmf{dbl_plain_arrow,lab.side=left,label=$\pi$}{p2,a}
 	      \fmf{dbl_plain_arrow,lab.side=left,label=$\omega$}{p1,b}

	      \fmfforce{220,90}{p2}
	      \fmfforce{220,10}{p1}

	      \fmfforce{270,90}{a}
	      \fmfforce{270,10}{b}

	      \fmfforce{0,90}{fb}
	      \fmfforce{0,10}{f}

      \end{fmfgraph*}

\end{fmffile}
\caption{Interaction with intermediate vector mesons.}
\label{fig2}
\end{center}
\end{figure}

The amplitude of the decay is described by the Feynman diagrams given in Figs.~\ref{fig1} and~\ref{fig2}. 
These diagrams are similar to diagrams used in Ref.~\cite{Arbuzov:2010xi} to describe the process
$e^++e^-\to\omega+\pi^0$.

The Lagrangian of quark-meson interactions in the framework of the extended NJL model was given 
in Refs.~\cite{VolkovEbertNagy,Arbuzov:2010xi,Arbuzov:2011zz}. 
Therefore, in present work we give only the expression for the amplitude describing the 
decay $\tau\to\pi\omega\nu$: 
\begin{equation}
T = G_F|V_{ud}| \bar\nu(1-\gamma^{5})\gamma^\mu\tau (T_{W\rho} + T_{\rho'})\epsilon_{\mu\nu\rho\sigma}p^{\rho}_{\omega}p^{\sigma}_{\pi}\omega^{\nu}\pi\,,
\end{equation}
where $G_F = 1.16637 \cdot 10^{-11} \: \mathrm{MeV^{-2}}$ is the Fermi coupling constant; 
$|V_{ud}| = 0.97428$ is the cosine of the Cabibbo angle, 
$p_\omega$ and $p_\pi$ are the $\omega$ and $\pi$ meson momenta.

The $T_{W\rho}$ term corresponds to the contribution given by the contact diagram and the diagram with 
an intermediate $\rho(770)$ meson. Using the factor for $W-\rho$ transition, we can get the expression that
coincides with one given by the vector meson dominance model:
\begin{equation}
T_{W\rho} = \frac{C_\rho}{g_{\rho_1}}\frac{1 - i\Gamma_\rho/m_\rho}{m_\rho^2 - p^2 - im_\rho\Gamma_\rho}m_\rho^2\,,
\end{equation}
where $m_\rho = 775.49$ MeV is the mass of $\rho(770)$ meson and $\Gamma(m_\rho^2) = 149.1\: \mathrm{MeV}$ 
is its total decay width.
The contribution of the amplitude with an intermediate $\rho(1450)$ meson reads
\begin{equation}
T_{\rho'} = C_{\rho'}C_{W\rho'}\frac{p^2}{m_{\rho'}^2 - p^2 - i\sqrt{p^2}\Gamma_{\rho'}(p^2)}\,.
\end{equation}
where $C_{W\rho'}$ corresponds to the $W-\rho'$ transition, it was defined in~\cite{Volkov:2012uh}.
The vertex constants $C_{\rho}$ and $C_{\rho'}$ are defined from the extended NJL model 
Lagrangian~\cite{Arbuzov:2010xi}:
\begin{equation}
\frac{C_{\rho}}{g_{\pi_1}} = \left(g_{\rho_1}\frac{\sin(\beta+\beta_0)}{\sin(2\beta_0)}\right)^2I_3 + \left(g_{\rho_2}\frac{\sin(\beta-\beta_0)}{\sin(2\beta_0)}\right)^2I_3^{ff} + 2g_{\rho_1}g_{\rho_2}\frac{\sin(\beta+\beta_0)}{\sin(2\beta_0)}\frac{\sin(\beta-\beta_0)}{\sin(\beta_0)}I_3^{f}\,,
\end{equation}
\begin{eqnarray}
-\frac{C_{\rho'}}{g_{\pi_1}} &=& g_{\rho_1}\frac{\sin(\beta+\beta_0)}{\sin(2\beta_0)}g_{\rho_1}\frac{\cos(\beta+\beta_0)}{\sin(2\beta_0)}I_3
\\ &+& g_{\rho_2}\frac{\sin(\beta-\beta_0)}{\sin(2\beta_0)}g_{\rho_2}\frac{\cos(\beta-\beta_0)}{\sin(2\beta_0)}I_3^{ff} \nn
\\ &+& g_{\rho_1}\frac{\sin(\beta+\beta_0)}{\sin(2\beta_0)}g_{\rho_2}\frac{\cos(\beta-\beta_0)}{\sin(2\beta_0)}I_3^{f} \nn
\\ &+& g_{\rho_2}\frac{\cos(\beta+\beta_0)}{\sin(2\beta_0)}g_{\rho_1}\frac{\sin(\beta-\beta_0)}{\sin(2\beta_0)}I_3^{f} \nn \,,
\end{eqnarray}
where $g_{\rho_1} = 6.14$, $g_{\rho_2} = 10.56$, $g_{\pi_1} = g_{\rho_1}/\sqrt{6}$, $\beta_0 = 61.44^{\circ}$, 
$\beta = 79.85^{\circ}$\footnote{In this work we use re-calculated values for the set of parameters of the extended 
NJL model which are a bit different from the ones used in the previous works.}
The definitions of integrals $I_3, I_3^{f}, I_3^{ff}$ was given in 
Ref.~\cite{VolkovEbertNagy}\footnote{In present work integrals $I_3,I_3^{f},I_3^{ff}$ was calculated 
in the $p^2$ approximation as it was done in more recent works~\cite{Arbuzov:2011zz,Arbuzov:2011rb} 
with 3-dimensional cut-off.}
Using the same set of parameters and the approach to loop integral computation, we re-calculated
the energy dependence of the $e^{+}+e^{-} \to \omega+\pi^0$ process cross-section. Here we get
a better description of the experimental data with respect to our earlier result~\cite{Arbuzov:2010xi}
obtained within the same model (but with another treatment of the loop integrals and other parameter
values). 
It allows us to get a better agreement with experimental data 
for $e^{+}e^{-} \to \pi\omega$ cross-section, see Fig.~\ref{eepiomega}.


\begin{figure}[ht]
\center{\includegraphics[width=0.7\linewidth]{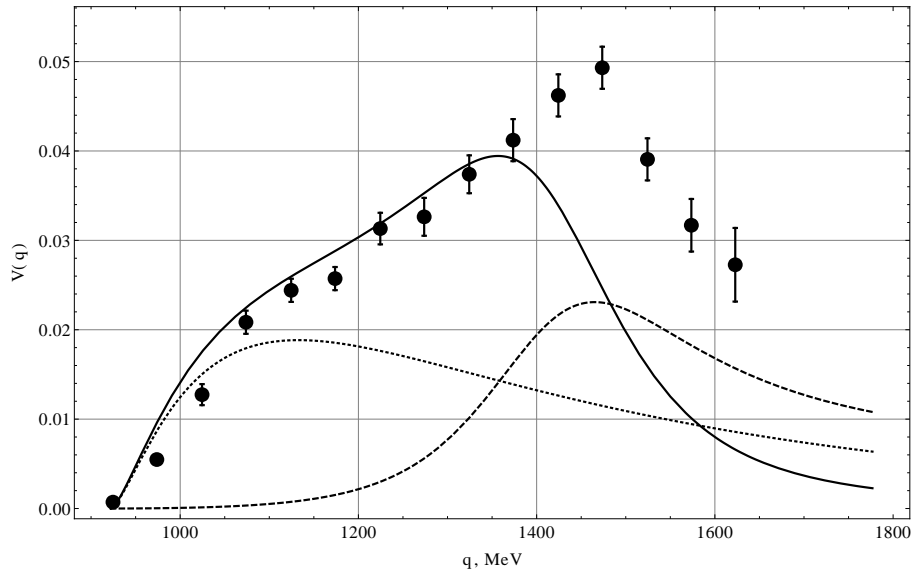}}
\caption{Comparison of CLEO~\cite{Edwards:1999fj} spectral function (dots) with the NJL predictions.
Solid, dotted and dashed lines are the total, $T_{W\rho}$ and $T_{\rho'}$ contributions, respectively.}
\label{tauomegepi}
\end{figure}

Use this formulas we get values for the branching of the $\tau\to\pi\omega\nu$:
\begin{equation}
	\mathrm{Br}^{NJL} = 1.85\%
\end{equation}

In experimental work~\cite{Edwards:1999fj} the results of the fit CVC model is given. 
With the help of the formula given in~\cite{Edwards:1999fj} we can get a prediction for the spectral 
density, see Fig.~\ref{tauomegepi}. The position of the peak in the NJL prediction for the spectral
density differs from the one seen in the data. First of all, this is because we took the standard 
value of the $m_{\rho'}=1465$~MeV value, while the fit of the CLEO data gives about $1520$~MeV.
Moreover, we did not take into account the contribution of the $\rho''$ intermediate state. The latter
is important for the spectral function at large $q$ values, while it is very much suppressed in the decay spectrum.

From our model one can get parameters for the density and the form factor: 
\begin{equation}
g_{\rho\omega\pi} = \frac{3g_{\rho}^2}{8\pi^2F_{\pi}} = 15.4~{\mathrm{GeV}}^{-1},
\end{equation}
\begin{equation}
A_1 = \frac{C_{\rho'}C_{W\rho'}}{C_{\rho}} = -0.13.  
\end{equation}
The comparison of experimental and theoretical values for these parameters 
is presented in Table~\ref{tbl1}.

\begin{table}[ht]
\begin{center}
\caption{Experimental and theoretical data.}
{\begin{tabular}{@{}llll@{}} 
\hline
 & $g_{\rho\omega\pi}$, 1/GeV & $A_1$ & $\mathrm{Br}(\tau\to\pi\omega\nu)$ , \% \\
\hline
 Theory & $15.40$ & $-0.13$ & $1.85 $ \\
 CLEO~\cite{Edwards:1999fj} & $16.10 \pm 0.06$ & $-0.23 \pm 0.02$ & $1.95 \pm 0.08$ \\ 
 ALEPH~\cite{Buskulic:1996qs} & -- & --  & $1.91 \pm 0.13$ \\
 SND-2011~\cite{Achasov:2012zz} & $15.75 \pm 0.45$ & $-0.29 \pm 0.09$ & -- \\
\hline
\end{tabular}}
\label{tbl1}
\end{center}
\end{table}


\begin{figure}[ht]
\center{\includegraphics[width=0.7\linewidth]{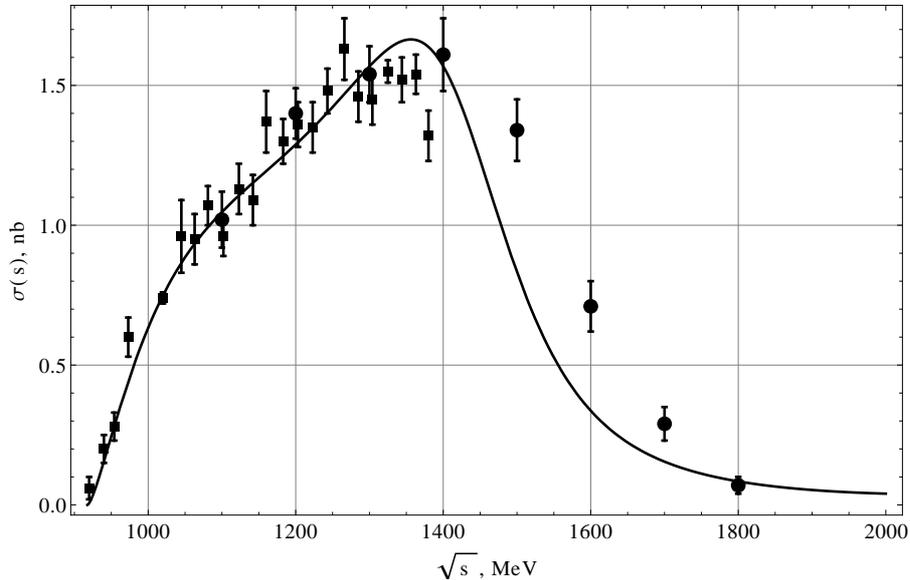}}
\caption{Comparison of experimental data of SND-2 
(squares~\cite{Achasov:2000wy} and dots~\cite{Achasov:2012zz}) for $e^{+}e^{-} \to \pi\omega$ 
with the NJL prediction (solid line).}
\label{eepiomega}
\end{figure}

\section{Conclusions}
The presented calculation show that the extended NJL model allows to describe the branching 
of the decay $\tau\to\pi\omega\nu$ in a satisfactory agreement with experimental data
without introduction of any additional arbitrary parameters.
This fact distinguishes our model from phenomenological approaches used earlier in 
Refs.~\cite{LopezCastro:1996xh,FloresTlalpa:2007bt,Guo:2008sh}.

We note that CLEO experimental~\cite{Edwards:1999fj} values for energy range from $1.4$ to $1.5$~GeV 
don't coincide with number of other experiments~\cite{Achasov:2012zz} and NJL prediction. 
It may affect to values for $\rho'$ mass and $\rho''$ contribution given by fits~\cite{Edwards:1999fj}.

In future works we are going to describe within the same model the $tau$ lepton decays with 
the creation of $\eta, \eta'$ mesons.

\section*{Acknowledgments}
We are grateful to E.~A.~Kuraev for useful discussions. 
This work was supported by the RFBR grant 10-02-01295-a.

\end{document}